\documentclass[
final            
  ]{aipproc}
\usepackage[OT1]{fontenc}
\usepackage[latin1]{inputenc}
\layoutstyle{8x11single}
\usepackage{epsfig} 
\usepackage{graphicx}
\usepackage{amssymb, amsmath, amsfonts}
\usepackage{epstopdf}
\usepackage{dcolumn}
\usepackage{bm}
\begin{document}

\def\vri{\vec{r}_{i}}
\def\vrj{\vec{r}_{j}}
\def\rij{r_{ij}}
\def\vrij{\vec{r}_{ij}}
\def\drij{\hat{r}_{ij}}
\def\vdr{\delta\vec{r}}
\def\dr{\delta{r}}
\def\s{\hat{s}}

\title{Theory of random packings}


\author{Chaoming Song, Ping Wang, Hern\'an A. Makse}{ address={ Levich
    Institute and Physics Department, City College of New York, New
    York, NY 10031, US}}


\begin{abstract}
  We review a recently proposed theory of random packings.  We
  describe the volume fluctuations in jammed matter through a volume
  function, amenable to analytical and numerical calculations.  We
  combine an extended statistical mechanics approach 'a la Edwards'
  (where the role traditionally played by the energy and temperature
  in thermal systems is substituted by the volume and compactivity)
  with a constraint on mechanical stability imposed by the isostatic
  condition.  We show how such approaches can bring results that can
  be compared to experiments and allow for an exploitation of the
  statistical mechanics framework.  The key result is the use of a
  relation between the local Voronoi volume of the constituent grains
  and the number of neighbors in contact that permits a simple
  combination of the two approaches to develop a theory of random
  packings. We predict the density of random loose packing (RLP) and
  random close packing (RCP) in close agreement with experiments and
  develop a phase diagram of jammed matter that provides a unifying
  view of the disordered hard sphere packing problem and further
  shedding light on a diverse spectrum of data, including the RLP
  state.  Theoretical results are well reproduced by numerical
  simulations that confirm the essential role played by friction in
  determining both the RLP and RCP limits. Finally we present an
  extended discussion on the existence of geometrical and mechanical
  coordination numbers and how to measure both quantities in
  experiments and computer simulations.
\end{abstract}

\maketitle


\section{I. Statistical mechanics of jammed matter}

Conventional Statistical Mechanics uses the ergodic hypothesis to
derive the microcanonical and canonical ensembles, based on the
quantities conserved, typically the energy $E$
\cite{landau-stat-mech}.  Thus the entropy in the microcanonical
ensemble is $S(E) = k_B \log \int \delta (E-{\cal H}(p,q)) dp dq$,
where ${\cal H}(p,q)$ is the Hamiltonian.  This becomes the canonical
ensemble with $\exp [- {\cal H} (\partial S/\partial E)]$. Experiments
\cite{chicago1,bideau,sdr,swinney} indicate that systematically shaken
granular materials show reversible behavior, and the analogue of the
conserved quantity is the volume $V$,
thus the micro-canonical ensemble or V-ensemble is
\cite{edwards2,blumenfeld,dauchot1,ciamarra,mk,E3,E6}:
\begin{equation}
  \Omega(V) = \exp[S(V)/\lambda] = \int \delta (V-{\cal W}(\vec{r_i})) 
\,\, \Theta_{\rm jam}(\vec{r_i})  \,\, d\vec{r_i},
   \label{eq}
\end{equation}
where $\Theta_{\rm jam}(\vec{r_i})$ is a function that defines the
jammed configuration.  As a minimum requirement the jamming function
$\Theta_{\rm jam}(\vec{r_i})$ should ensure touching grains, and
obedience to Newton's force laws. $\vec{r_i}$ denotes the particle
positions in the system and ${\cal W}(\vec{r_i})$ is the volume
function defining the volume associated with each grain (see
below). This gives a canonical ensemble of $\exp [-{\cal W} (\partial
S/\partial V)]$.
 Just as $\partial E/\partial S = T$ is the temperature in
equilibrium system, the temperature-like variable in granular systems
is the compactivity $X=\partial V/\partial S$.
In Eq. (\ref{eq}), $\lambda$ is the analogue of the Boltzmann
constant.

Thermodynamic analogies may illuminate methods for attempting to solve
certain problems, but inevitably fail at some point in their
application. The mode of this failure is an interesting phenomenon,
illustrated by the compaction experiments of the groups of Chicago,
Texas, Paris and Schlumberger \cite{chicago1,bideau,sdr,swinney}. They
have shown that reversible states exist along a branch of compaction
curve where statistical mechanics is more likely to work.  Conversely,
experiments also showed a branch of irreversibility where the
statistical framework is not expected to work.  Poorly consolidated
formations, such as a sandpile, are irreversible and a new
``out-of-equilibrium'' theory is required to describe them.  Below we
focus on a theoretical description of the reversible branch of the
compaction curve focusing on a theory of the random close packed
state.

The canonical partition function in the V-ensemble is the starting
point of the statistical analysis of jamming:
\begin{equation}
\mathcal{Z}(X) =
\int g({\cal W}) \,\, e^{-{\cal W}/X} \,\, \Theta_{\rm jam} \,\, d{\cal W},
\label{Q}
\end{equation}
where $g({\cal W})$ is the density of states for a given volume $\cal
W$.

From Eq. (\ref{Q}) we identify three minimal steps in developing
analytical solutions which are discussed in the next sections. Section II
 discusses the need for a volume function in
terms of the contact network.  Section III discusses the
need for a proper definition of jammed state that allows one to define
$\Theta_{\rm jam}$. Section IV discusses the density of
states.
Finally in Section V we explain the geometrical and mechanical
coordination numbers and how to measure them in Section V.A, and we
conclude in Section VI.


\section{II. Volume Function}
\label{volume-function}

While it is always possible to quantify the total volume of the
system, it is unclear how to treat the volume fluctuations at the
grain level.  The first step to study the V-ensemble is to find the
volume ${\cal W}(\vec{r_i})$ associated to each particle $\vec{r_i}$
that successfully tiles the system.  This is analogous to the additive
property of energy in equilibrium statistical mechanics.

Initial attempts included a model volume function under mean-field
approximation \cite{edwards2}, the work of Ball and Blumenfeld
\cite{ball} and simpler versions in terms of the first coordination
shell \cite{E8}. These definitions are problematic since some are not
additive, others present problems in polydisperse systems or are
proportional to coordination contrary to expectation. In
Ref. \cite{jamming1,jamming2} we have found an analytical form of the
volume function in three-dimensions and demonstrated that it is the
Voronoi volume of a particle $i$:

\begin{equation} \label{vor1}
{\cal W}_i^{\rm vor} = \frac{1}{3} \oint \left(\min_{\s\cdot\drij >
  0}(\frac{\rij}{2\s\cdot\drij})\right)^3 ds,
\end{equation}
where $\vrij$ is the vector from the position of particle $i$ to that
of particle $j$, the integration is done over all the directions $\s$
forming an angle $\theta_{ij}$ with $\vrij$ as in Fig. \ref{phase}a,
and $R$ is the radius of the grain.  $R$ will be set to unity for
simplicity.  While this formula may seem complicated, it has a simple
interpretation depicted in Fig. \ref{phase}a.

The Voronoi construction is additive and successfully tiles the total
volume. Prior to this result, there was no analytical formula to
calculate the Voronoi volume in terms of the contact network $\rij$.
A further simplification arises when we consider isotropic systems.
Then the volume function reduces to the orientation volume, without
the average over $\s$. We define the reduced free orientational volume
function as
\begin{equation}
w^s \equiv \frac{{\cal W}_i^s - V_g}{V_g},
\label{vor2}
\end{equation}
with ${\cal W}_i^s \equiv V_g \left(\frac{1}{2R} \min_{\s\cdot\drij >
    0} \frac{\rij}{\s\cdot\drij}\right) ^ 3$, see Fig. \ref{phase}a
($V_g$ is the particle volume).  This equation allows theoretical
analysis in the V-ensemble since it reduces the complicated definition
(\ref{vor1}) to a more amenable ``one-dimensional'' volume which can
be treated analytically.

The next step is to develop a theory of volume fluctuations to coarse
grain ${\cal W}_i^s$ over a mesoscopic length scale.  We call this the
quasi-particle approximation. It could be considered as well as a
mean-field approximation, although mean field is supposed to be exact
in infinite dimensions. The approximations used in the present theory
are supposed to get better as the dimension increases, but we cannot
claim that the theory is exact in large dimensions. Thus, we prefer to
call our approximation ``quasi-particle'' in the spirit of Landau and
the quasiparticles as ``coordinons''.

The coarsening reduces the degrees of freedom to one variable, the
coordination number of each grain, and defines an average volume
function which is more amenable to statistical calculations than
Eq. (\ref{vor1}) as shown in \cite{jamming2}:
\begin{equation}
w(z) = \langle w^s\rangle_i =  \frac{2\sqrt{3}}{z},
\label{hamil}
\end{equation}
valid for monodisperse hard spheres where $z$ is the geometrical
coordination number. For now on we assume $V_g=1$ for simplicity. The
available volume per grain is inversely proportional with the
coordination number, in agreement with the X-ray tomography
experiments (see Fig. 6 in \cite{aste} where the volume fraction is
$\phi^{-1}= w + 1$).


\section {III. Definition of jamming via $\Theta_{\rm jam}$: Isostatic
  ensemble}
\label{definition}

The definition of the constraint function $\Theta_{\rm jam}$ is
intimately related to the proper definition of a jammed state , with a
minimum requirement of mechanical equilibrium. In an attempt to define
the jammed states in a rigorous mathematical way, Torquato and
coworkers have proposed three categories of jamming \cite{salvatore1}:
locally jammed, collectively jammed and strictly jammed based on
geometrical constraints.  Unfortunately this definition cannot be
easily extended to frictional systems since it is based on geometry
and does not include the contact forces. Other approaches based on
minima of the potential energy landscape also fail since such a
potential does not exist for frictional grains due to their
path-dependency.

Then we define an alternative approach to characterize jamming for the
general case of frictional granular matter.  In \cite{jamming2} we
propose the isostatic condition \cite{alexander,mouk1} as a possible
formulation of jamming.  The isostatic condition implies a mechanical
coordination number to be $Z=2d$, where $d$ is the dimension, for
frictionless spherical particles and $Z=d+1$ for infinitely rough
particles (with interparticle friction coefficient $\mu\to \infty$).
Numerical simulations \cite{makse,ohern}, experiments \cite{E21} and
theoretical work \cite{mouk1} suggest that at the jamming transition
the system becomes exactly isostatic. However, no rigorous proof of
this statement exist.  It should be noted that $Z=6$ is a necessary
but not sufficient condition for isostaticity \cite{mouk1}. A
kinematic condition has to be satisfied as well, which refers to the
location of the center of the particles that are determined only by
the length of the vectors that join the center of contacting
particles. In other words the equations of equilibrium needs to be
independent of each other.  While this condition has been proved to
exist in frictionless packings \cite{mouk1}, the problem of frictional
packings, even with infinite friction, remains open. Thus, to be
precise, the only point where we can claim isostaticity is the
frictionless point. In any case, we consider that we are able to
extend the isostatic condition to infinite frictional packings as well
at $Z=4$.  Interpolating between the two limits, there exist packings
of finite $\mu$; the coordination number smoothly varies between
$Z(\mu=0)=6$ and $Z(\mu\to \infty)\to 4$ \cite{silbert,jamming2}. 

We notice that while the relationship between friction and $Z$ may not
be unique, the theory is only based on $Z$.  Thus, given the
mechanical coordination number we predict the state of the packing
with the compactivity. Additionally, we conjecture that $\mu$
determines $Z(\mu)$ if we follow certain protocols as discussed in
\cite{jamming2}. These protocols imply compression of a packing from an
unjammed state to jamming by following one single continuous path.
That is, the path dependent shear forces are not reset by a sudden
change in the preparation path.  It should be stated that there are
other protocols, like shear cycling, that start with a given $Z$ and
can produce packings that continuously compactify until RCP (and even
beyond). This is done by effectively changing the path followed by the
shear forces at every cycle. Thus, for shear cycling protocols, a
unique relation may not be expected between $Z$ and $\mu$.  However,
the theoretical results as derived below are still valid in shear
cycling experiment since they concern the relation between $Z$ and
volume. In fact, a shear experiment may be the easiest way to obtain
the packings at the RLP line as explained below.

Assuming that a system of hard spheres is isostatic at the jamming
transition, Eq. (\ref{Q}) can be written in terms of $z$ and the
mean-field Eq. (\ref{hamil}) can be used in the single-particle (or
more precisely quasi-particle) partition function:
\begin{equation}
{\mathcal Z}_{\rm iso}(X) = \int_Z^6 e^{-w(z)/X} g(z) dz.
\label{partition}
\end{equation}
We notice that in principle the coordination number of each particle
takes only integer values. However, the coordination number in
(\ref{partition}) implies a coarse graining over several particles,
and therefore can take non-integer values. Thus the use of an integral
in (\ref{partition}) instead of a sum is justified

\begin{figure*} \centering { \hbox { (a)
\resizebox{5cm}{!}{\includegraphics{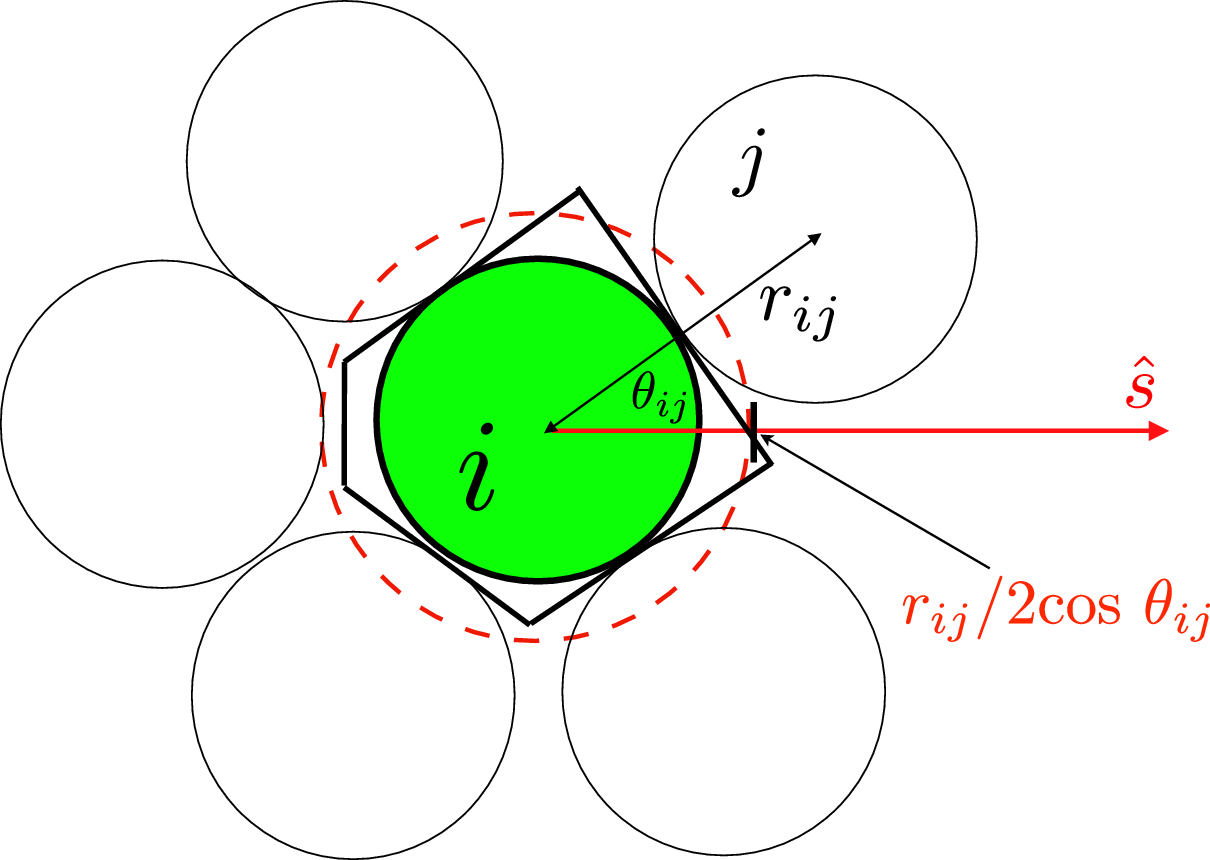}}
(b) \resizebox{5cm}{!}{\includegraphics{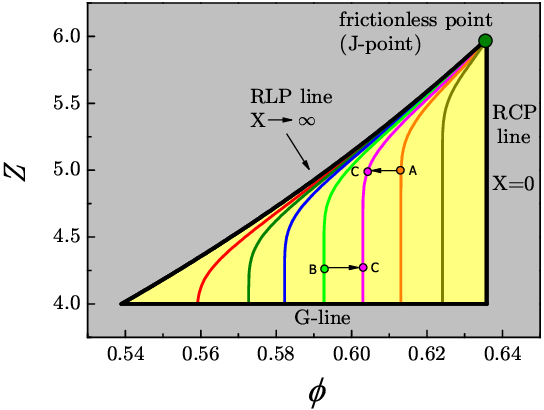}} (c)
\resizebox{5cm}{!}{\includegraphics{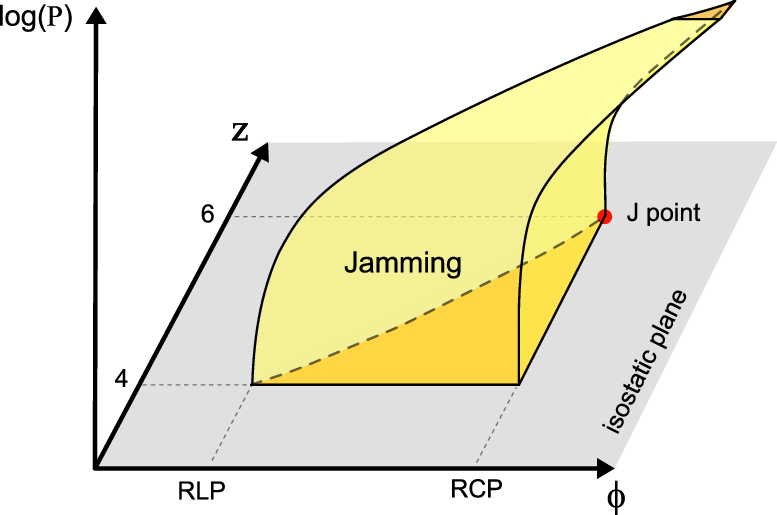}} } } 
\vspace{.5cm}
\caption{ (a) Schematics of the Voronoi volume and the orientational
  volume associated with particle $i$.  The boundary of the Voronoi
  cell (shown in two-dimensions for simplicity) corresponds to the
  irregular pentagon in black which defines ${\cal W}_i^{\rm
    vor}$. The limit of the Voronoi cell of particle $i$ in the
  direction $\s$ is the minimum of $r_{ij}/2 \cos \theta_{ij}$ over
  all the particles in the packing, as indicated. This defines the
  orientational volume ${\cal W}_i^s$ which is the volume of the
  sphere of radius $r_{ij}/2 \cos \theta_{ij}$ defined by the dash red
  circle in the figure.  (b) Phase diagram of jamming in the hard
  sphere plane under the isostatic assumption.  All the disordered
  packings fall within the yellow triangle demarcated by the RCP and
  RLP lines and the G-line. The isocompactivity lines are in color.
  (c) Generalization of the phase diagram to the space $(Z,\phi,p)$.}
\label{phase}
\end{figure*}

This ensemble is referred to as the {\it Isostatic-ensemble}. Note
that the upper limit of integration is $z=6$ \cite{jamming2}. This
implies that only disordered packings are included. The solution of
such a partition function for monodisperse hard spheres has been done
in \cite{jamming2} revealing the phase diagram depicted in
Fig. \ref{phase}b.

This phase diagram predicts a series of important results, such as the
value of RCP at $X = 0$,

\begin{equation}
\phi_{\rm RCP}=6/(6+2\sqrt{3}),
\end{equation}
and the
lowest density of the RLP at $X = \infty$, 
\begin{equation}
\phi_{\rm
RLP}=4/(4+2\sqrt{3}),
\end{equation}
in close agreement with experiments.
The diagram restricts the possible packings to the yellow triangle in
Fig. \ref{phase}b, ranging from frictionless systems with $Z=6$, to
infinitely rough grains in the $Z=4$ granular line or G-line.


\section{IV. Density of states}
\label{density}

A difficult problem is the determination of the density of states
$g({\cal W})$ in Eq. (\ref{Q}).  For the simplest case of the
Iso-ensemble from Eq (\ref{partition}), the density of states reduces
to 
\begin{equation}
g(z)=(h_z)^{z},
\label{g}
\end{equation}
where $h_z$ is a small microscopic constant arising due to the
discrete volume space of configurations \cite{jamming2}.  The
situation is analogous to the discreteness of the configuration space
imposed by the Heisenberg uncertainty principle in quantum mechanics.
The formula is analogous to the factor $h^{-d}$ for the density of
states in equilibrium statistical mechanics.  While the degrees of
freedom $\{ p_i,q_i \}$ are continuous, the uncertainty principle
imposes the discreteness $(\Delta p, \Delta q)$ in the configurational
space given by $\Delta p \Delta q \sim h$.  This consideration allows
for the approximate solution explained in the above section and
depicted in the phase diagram of Fig.  \ref{phase}b.

\section{V. Geometrical versus mechanical coordination number}
\label{geometrical}

It is important to note that the derivation of the volume function in
Section II implies nothing about the value of the
contact forces; the volume function represents the contribution
arising purely from the geometry of the packing.  Thus, the
coordination number $z$ appearing in Eq. (\ref{hamil}) is the {\it
  geometrical} coordination number related to volume, which is
different from the {\it mechanical} coordination number $Z$ that
counts the number of contacts per particle with non-zero force related
to the isostatic condition and force network.

Having acknowledged a difference between the geometrical coordination
number $z$ in Eq. (\ref{hamil}) and the mechanical coordination number
$Z$ which counts only the contacts with non-zero forces, below we
discuss the bounds of $z$ and how to measure it.

Since some geometrical contacts may carry no force, then we have:

\begin{equation}
  Z\le z. \label{coord_low}
\end{equation}
To show this, imagine a packing of infinitely rough ($\mu\to
\infty$) spheres with volume fraction close to $0.64$. There must
be $z=6$ nearest neighbors around each particle on the average.
However, the mechanical balance law requires only $Z=4$ contacts
per particle on average, implying that $2$ contacts have zero
force and do not contribute to the contact force network.

\begin{figure}
  \centering {
      \resizebox{6cm}{!}{\includegraphics{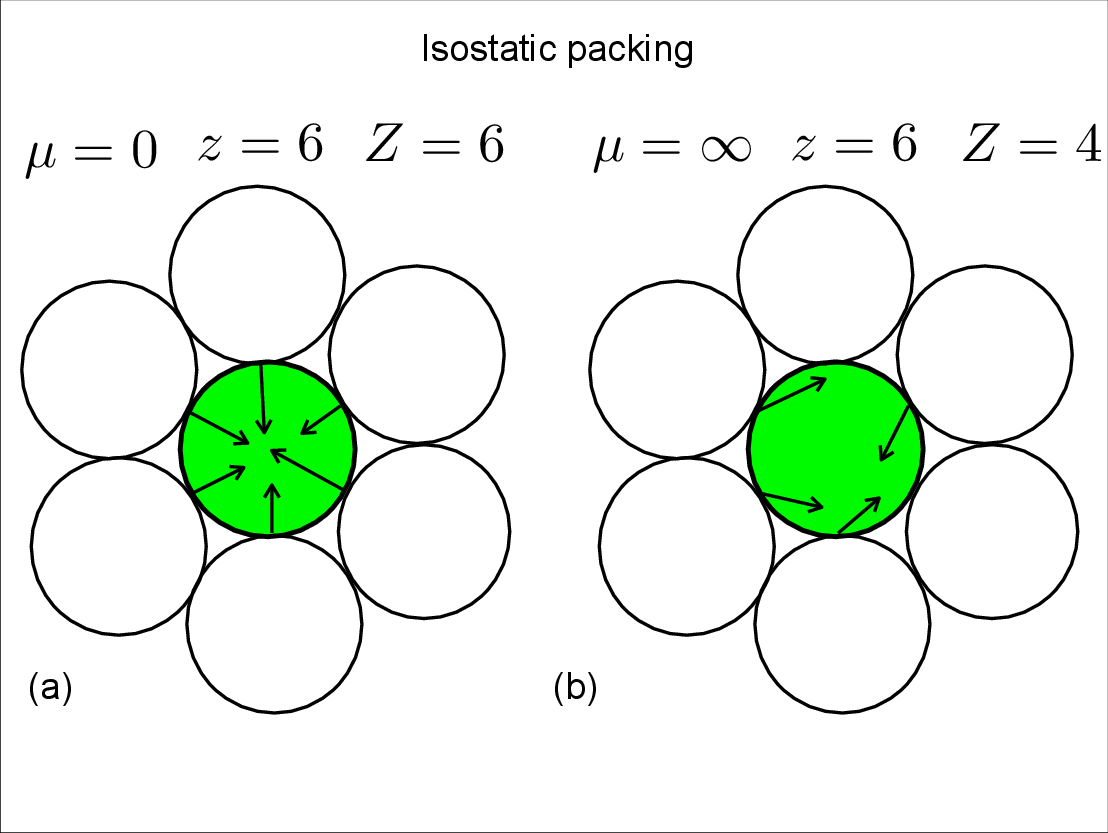}}
    }
\caption{ (a) Consider a frictionless packing at the isostatic
    limit with $z=6$.  In this case the isostatic condition implies
    also $Z=6$ mechanical forces from the surrounding particles. (b)
    If we now switch on the tangential forces using the same packing
    as in (a) by setting $\mu\to\infty$, the particle requires only
    $Z=4$ contacts to be rigid. Such a solution is guaranteed by the
    isostatic condition for $\mu\to\infty$. Thus, the particle still
    have $z=6$ geometrical neighbors but only $Z=4$ mechanical ones. }
\label{possible}
\end{figure}

Such a situation is possible as shown in Fig. \ref{possible}: starting
with the contact network of an isostatic packing of frictionless
spheres having $Z=6$ and all contacts carrying forces (then $z=6$ also
as shown in Fig. \ref{possible}a), we simply allow the existence of
tangential forces between the particles and switch the friction
coefficient to infinity. Subsequently, we solve the force and torque
balance equations again for this modified packing of infinitely rough
spheres but same geometrical network, as shown in Fig. \ref{possible}b
[Notice that the shear force is composed of an elastic Mindlin
component plus the Coulomb condition determined by $\mu$.  Thus when
$\mu \to \infty$, the elastic Mindlin component still remains].

The resulting packing is mechanically stable and is obtained by
setting to zero the forces of two contacts per ball, on average,
to satisfy the new force and torque balance condition for the
additional tangential force at the contact.  Such a solution is
guaranteed to exist due to the isostatic condition: at $Z=4$ the
number of equations equals the number of force variables.  Despite
mechanical equilibrium, giving $Z=4$, there are still $z=6$
geometrical contacts contributing to the volume function.

Therefore, we identify two types of coordination number: the
geometrical coordination number, $z$, contributing to the volume
function and the mechanical coordination number, $Z$, measuring
the contacts that carry forces only.  This distinction is crucial
to understand the sum over the states and the bounds in the
partition function.

We have established a lower bound of the geometrical coordination in
Eq. (\ref{coord_low}). The upper bound arises from considering the
constraints in the positions of the rigid hard spheres.  For hard
spheres, the $Nd$ positions of the particles are constrained by the
$Nz/2$ geometrical constraints, $|\vec{r}_{ij}| = 2R$, of
rigidity. Here $d$ is the dimension. Thus, the number of contacts
satisfies $Nz/2\le Nd$, and $z$ is bounded by:

\begin{equation}
  z\le 2d. \label{coord1}
\end{equation}
Notice that this upper bound applies to the geometrical
coordination, $z$ and not to the mechanical one, $Z$, and it is
valid for any system irrespective of the friction coefficient,
from $\mu = 0 \to \infty$.

In conclusion, the mechanical coordination number, $Z$, ranges
from 4 to 6 as a function of $\mu$, and provides a lower bound to
the geometrical coordination number, while the upper bound is
$2d$.  A granular system is specified by the interparticle
friction which determines the average mechanical coordination at
which the system is equilibrated, $Z(\mu)$. The possible
microstates in the ensemble available for this system follow a
Boltzmann distribution Eq. (\ref{Q}) for states satisfying the
following bounds:
\begin{equation}
\label{bounds-eq} Z(\mu) \le z \le 2d = 6.
\end{equation}

\begin{figure}
\centering \resizebox{9cm}{!}{\includegraphics{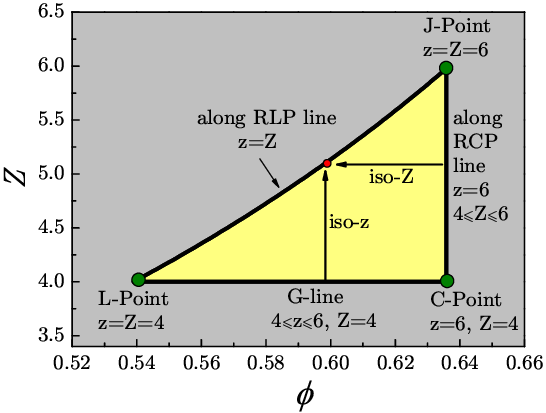}}
\caption{Summary of the theoretical findings regarding the range of
  $z$ and $Z$ along the different iso-$z$, iso-$Z$, and iso-$X$ lines
  and the J, C, and L-points in the phase diagram. } \label{z-summary}
\end{figure}

\subsection{V.A. How to measure the geometrical coordination number}

Measuring the geometrical coordination number can be a tricky task, in
principle.  At the onset, it is the coordination number of a single
quasiparticle.  What we measure in a real packing (numerically or
experimentally generated) is an ensemble average of many quasiparticle
according to the partition function Eq. (\ref{partition}). Thus,
rigorously speaking, it is not possible to isolate a quasi-particle
and measure its properties in a real packing. Beyond this caustic and
somehow pessimistic remark, yet rigorous, below we offer light at the
end of the tunnel by using the theoretical predictions to define an
approximative, yet accurate, way to measure the geometrical
coordination.

Figure \ref{z-summary} summarizes the predictions of the theory
regarding the behaviour of $z$ and $Z$ for the packings in the phase
diagram. First, we can think that the quasiparticle behaviour is
revealed when the system has infinite compactivity and behaves like a
non-interacting ``gas'' of quasiparticles. This is the behaviour found
along the RLP line in the phase diagram of Fig. \ref{phase}b. Indeed,
along this line, the packings have the highest entropy \cite{jamming3}
and therefore they are the most likely to be found numerically and
experimentally. Indeed, this is what is found in our numerical
simulations \cite{jamming2,jamming3}. Along this line we expect that
the mechanical and geometrical coordinations are the same, $z\approx
Z$.  This result comes about since at infinite compactivity we are
exploring the states with the highest volume or lower volume fraction.
In the partition function Eq. (\ref{partition}), these are states with
$z\approx Z$. It should be noticed that the density of states plays a
role in this argument. We are assuming that the density decays very
rapidly owning to a small constant $h_z\to 0$ in Eq. (\ref{g}).  This
is of course a very reasonable assumption, since $h_z$ is related to
the discretization of the volume space of configurations, and it is
like a Planck constant of granular matter.  However, we may even relax
this consideration and allow this constant to be of the order 1.  The
above argument is still correct, but the only difference is that the
RLP limit is not at infinite compactivity but in the limit of
$X\to0^-$. That is, the highest volume corresponds to negative
temperature states, which are even hotter than $X\to +\infty$.  An
extended explanation of this point is given in \cite{jamming3} and its
follow up paper in cond-mat.

The second interesting aspect of the theory is the prediction that along
the RCP line in Fig. \ref{z-summary}, the geometrical coordination
number is constant and equal to the maximum coordination, which
produces the minimum Voronoi volume: $z=6$. At the same time, the
mechanical coordination number varies from 4 to 6 as we reduce
friction to zero.  This result is explained with the analysis offered
in Fig. \ref{possible} and explained above. The conclusion is that the
packings along the RCP line are geometrically the same (with $z=6$ all
of them) but they differ in the value of the forces between the
particles as $Z$ varies from 6 to 4.  This prediction can be also
applied to other packings with lower $\phi$. For a given volume
fraction, the packings along the iso-$z$ vertical line in
Fig. \ref{z-summary} should have approximately the same $z$ but
different $Z$ from 4 to the maximum $Z$ given by the RLP line.

Using this theoretical result, it is easy then to define the
geometrical coordination number for real packings and propose a clear
way to measure it. The idea is to inflate the particles
infinitesimally by a $\Delta r$ value and measure the contacting
particles. By setting $\Delta r=0$, we clearly measure the mechanical
coordination. By considering an infinitesimally small $\Delta r$ we
should measure the geometrical one.  The question is to know what
value of $\Delta r$ to use.  Here is where the theory comes handy.  We
know that along the RCP line the packings are the same geometrically.
So, whatever the definition of coordination number we use, it should
satisfy that after a given $\Delta r$, we should find the same packing
structure for the packings along the RCP line and below the given
$\Delta r$, the structure should change reflecting the different
values of $Z$ for different packings in the RCP line.

Using these considerations, we identify the geometrical coordination
as follows.  Two particles that may not be in contact (giving rise to
a zero force) may be close enough to be considered as contributing to
the geometrical coordination.  Indeed, it is known that the radial
distribution function $g(r)$ has a singularity, $g(r) \sim
(r-0.5)^{-0.5}$ \cite{silbert}, implying that there are many particles
almost touching.
We introduce a modified radial distribution function (RDF) $g_z(r)$ in
order to approximately identify $z$ and $Z$ from real packings:

\begin{equation}\label{dgz}
  g_z(r)=\frac{1}{N}\frac{R^2}{r^2}\sum^N_i\sum^N_{j\neq
    i}\Theta\Big(\frac{r_{ij}}{r-R}-1\Big)\Theta\Big(\frac{r+R}{r_{ij}}-1\Big), \ \ r > R
\end{equation}
where $R$ is the radius of particle, $N$ is the number of particles,
$r_{ij}$ is the distance of two particle's centers,
$r_{ij}=|\vec{r}_i-\vec{r}_j|$, and $\Theta$ is the Heaviside step
function. The RDF describes the average value of the number of grains
in contact with a virtual particle which has been inflated up to a
radius $r\ge R$, and the factor of $R^2/r^2$ is the ratio of a real
sphere's area and the virtual one's.  Without the normalization factor
$R^2/r^2$, Eq. (\ref{dgz}) is the same definition of coordination as
used by Torquato and Zamponi in their analysis of infinite pressure
jammed hard sphere glasses \cite{LS2,parisi}. This factor is not
crucial for our analysis (as the main constraint is that when $r=R$ we
should get $Z$) as explained below, but we argue that it is useful
since we need to proper normalize by the fact that we are inflating
the balls. Please notice that Eq. (29) in Supplementary Information
Section of \cite{jamming2} contains a typo. The correct definition is
Eq. (\ref{dgz}).

$g_z(r)$ measures the number of balls with their volume intersecting
the surface of a sphere of radius $r$ measured from the center of a
given ball.  When $r=R$ in (\ref{dgz}) we obtain the mechanical
coordination number while the geometrical one is obtained for a small
value $\Delta r = \frac{r-R}{2R} \ne 0$ for which we distinctly find a
signature from computer simulations, unambiguously defining it at
$\Delta r = 0.04$ for the system size used by following the packings
along the RCP line.

\begin{figure}
\centering \resizebox{6cm}{!}{\includegraphics{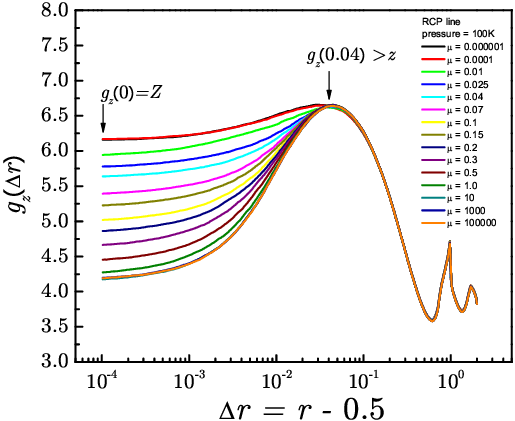}}
\caption{$g_z(\Delta r)$ of packings with various friction coefficient
  $\mu$ along RCP line. We set $2R=1$.} \label{gzr_RCP}
\end{figure}
\begin{figure}

\centering \resizebox{6cm}{!}{\includegraphics{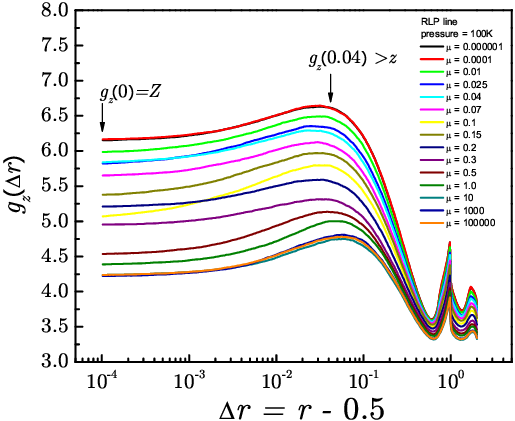}}
\caption{$g_z(\Delta r)$ of packings with various friction coefficient
$\mu$ along RLP line.} \label{gzr_RLP}
\end{figure}

Figures \ref{gzr_RCP} and \ref{gzr_RLP} plot the $g_z(\Delta r)$ of
packings with various friction coefficient $\mu$ along the RCP and RLP
lines respectively.  Following the definition of Eq. (\ref{dgz}),
$g_z$, with $\Delta r = 0$, should be directly equal to the mechanical
coordination number, $Z$, and should range from $4$ to $6$ along both
RCP and RLP (if $h_z\ll1$) lines which is confirmed by our numerical
simulations in Figs. \ref{gzr_RCP} and \ref{gzr_RLP}, respectively.

More importantly, as shown in the figures, we find that $g_z(\Delta
r)$ along the RCP line is exactly the same for all the packings when
$\Delta r>0.04$ as shown in Fig. \ref{gzr_RCP}.  For $\Delta r<0.04$,
Fig. \ref{gzr_RCP} shows that the packings have different mechanical
coordination numbers from 4 to 6.  This is exactly what the theory
predicted. All these packings have actually the same geometrical
structure evidenced when $\Delta r > 0.04$ but with different
mechanical coordinations which appears only in a difference observed
for $\Delta r<0.04$.  Based on this analysis, we then define the
geometrical coordination number as the one appearing at $\Delta r =
0.04$.  We identify the geometrical coordination number as
$z=g_z(0.04)$ under the accuracy of the simulations and for this
particular system size $N=10,000$ (we notice though that this value
may depend on system size).  

It is important to note that in terms of the radial distribution
function, $g(r)$, nothing really happens at $\Delta r =0.04$: that is,
there is no peak in $g(r)$ at 0.04 and the second peak after the first
coordination shell appears for larger $\Delta r$ in $g(r)$. On the
other hand, we clearly see a peak in Fig. \ref{gzr_RCP} at this
value. We point out that the peak at 0.04 is the byproduct of the
normalization factor $R^2/r^2$ in Eq. (\ref{dgz}). This factor is
suggested since we need to renormalize by the area of the virtual
sphere. However, other factors, for instance $R^3/r^3$ in
Eq. (\ref{dgz}) would produce a peak in Fig. \ref{gzr_RCP} located in
another position. The important fact is not the location of the peak,
but the fact that above 0.04 all the functions in Fig. \ref{gzr_RCP}
coincide.  This is the basis of the definition of 0.04 as the location
to define the geometrical coordination number.  At this position,
there is no peak in the $g(r)$. Indeed, the second peak in $g(r)$
beyond the first coordination shell appears much further around
$\Delta r \approx 1$ and are indeed also identified by $g_z(\Delta r)$
as can be seen in Fig. \ref{gzr_RCP} and Fig. \ref{gzr_RLP} as
well. Therefore the peak associated with the geometrical coordination
number is not revealed from the structure in $g(r)$. It has a more
subtle meaning as explained above.

It is also important to note that in experiments there is always an
uncertainty in measuring the position of the particles. According to
our analysis an small uncertainty of a few percent will render the
mechanical coordination into the geometrical one. Thus experiments
will be very difficult to differentiate between $z$ and $Z$. A
possible solution to this problem is to use complementary fluorescent
techniques \cite{E21,brujic2,brujic3} to obtained a signal when the
particles are carrying a force and not to rely on geometrical
reconstruction of index-matched images or X-ray tomography.  Another
route would be to obtain approximate coordinates from experiments and
then use them as input into a Molecular Dynamics simulations to obtain
the exact force balance for each particle.  This last approach may
provide the final way to accurately measure the coordinations of the
particles with accuracy. In our website http://www.jamlab.org we offer
the computer codes to perform MD simulations with Hertz-Mindlin forces
as well the code to calculate the entropy of the packings.

The theoretical analysis is also confirmed in the RLP packings.  Along
the RLP line, Fig. \ref{gzr_RLP}, we find that the geometrical
coordination number as extracted from $g_z(0.04)$ is very close to the
mechanical one.  Since the RLP line is at $X\to \infty$ and $h_z\ll
1$, all the states along RLP have $z\approx Z$ as we move along the
line varying the friction coefficient.  Thus, the numerical results
confirm the theory.

In conclusion, a prescription to measure the geometrical coordination
is the following: First we identify a theoretical way to define
it. For instance, here we use the theoretical prediction that all the
packings along RCP have the same $z$.  With the proper definition of
coordination number for an inflated particle, Eq. (\ref{dgz}), we
calculate the coordination as a function of $\Delta r$.  This
identifies $\Delta r = 0.04$ as the position to obtain the geometrical
$z$. We then explore any packing (not only at RCP line) and apply
Eq. (\ref{dgz}) at $\Delta r = 0.04$ and obtain $z$.  Figure
\ref{z-summary} summarizes the theoretical predictions of values of
$z$ and $Z$ for all the packings in the phase diagram.  If the reader
still has doubts about the difference between $z$ and $Z$ we offer a
final more vivid way to understand it in terms of the famous kissing
number conjecture from Newton and Gregory due to a remark of
A. Coniglio (private communication): a mechanical contact is like a
French kiss while a geometrical contact is any other inconsequential
kiss.

\section{VI. Conclusions}
\label{conclusions}

In conclusion, using Edwards statistical mechanics we have elucidated
some aspects of RLP and RCP in the disordered spherical packing
problem.  The phase diagram introduced here serves as a beginning to
understand how random packings fill space in three dimensions.  The
comparative advantage of the present approach over extensive work done
in the past, is in the classification of all packings through $X$, $Z$
and $\phi$ in the theoretical phase diagram from where these studies
could be systematically performed. This classification guides the
search for indications of jamming from a systematic point of view,
through the exploration of all jammed states from $\mu=0$ to $\mu\to
\infty$.  Our results not only apply to packings at the jamming
transition in the limit of hard spheres, but may also be extended to a
general phase diagram as sketched in Fig. \ref{phase}c to include
states with finite nonzero pressure. Such states are described by an
angoricity in addition to the compactivity as developed
here. Extensions to other dimensions, polydisperse systems and other
shapes of particles like ellipsoids and sphero-cylinders are being
worked out with the goal of developing a unifying thermodynamic view
of the physics of packings.

\begin{theacknowledgments} 
  This work is supported by the National Science Foundation, CMMT
  Division and the Department of Energy, Office of Basic Energy
  Sciences, Geosciences Division.
\end{theacknowledgments}

\bibliographystyle{aipproc}

\end{document}